\documentclass[aps,pre,twocolumn,showpacs,superscriptaddress,
groupedaddress]{revtex4-1}
\DeclareMathAlphabet{\mathpzc}{OT1}{pzc}{m}{it}
\usepackage{amsfonts}
\usepackage{graphicx}
\usepackage[symbol]{footmisc}

\usepackage[normalem]{ulem}

\input{defs}
\makeatletter
\let\@fnsymbol\@arabic
\makeatother
\begin{document}

\title{Experimental Realization of a Reconfigurable Electroacoustic Topological Insulator}

\author{Amir Darabi}
 \altaffiliation{Woodruff School of Mechanical Engineering, Georgia Institute of Technology, Atlanta, GA}
 \author{Manuel Collet}
 \altaffiliation{LTDS-UMR CNRS 5513, Ecole Centrale de Lyon (ECL), 36 Avenue Guy de Collongue, 69134 Ecully Cedex, France}
\author{Michael J. Leamy}  \altaffiliation{Woodruff School of Mechanical Engineering, Georgia Institute of Technology, Atlanta, GA}
\email{michael.leamy@me.gatech.edu}

\begin{abstract}
A substantial challenge in guiding elastic waves is the presence of reflection and scattering at sharp edges, defects, and disorder. Recently, mechanical topological insulators have sought to overcome this challenge by supporting back-scattering resistant wave transmission. In this Letter, we propose and experimentally demonstrate the first \emph{reconfigurable electroacoustic} topological insulator exhibiting an analog to the quantum valley Hall effect (QVHE). Using programmable switches, this phononic structure allows for rapid reconfiguration of domain walls and thus the ability to control back-scattering resistant wave propagation along dynamic interfaces for phonons lying in static and finite-frequency regimes. Accordingly, a graphene-like Polyactic Acid (PLA) layer serves as the host medium, equipped with periodically arranged and bonded piezoelectric patches, resulting in two Dirac cones at the $K-$points. The PZT patches are then connected to negative capacitance external circuits to break inversion symmetry and create nontrivial topologically-protected bandgaps. As such, topologically protected interface waves are demonstrated numerically and validated experimentally for different predefined trajectories over a broad frequency range. 
\end{abstract}
\maketitle

The need for lossless information carriers has generated increased interest in topologically-protected structures in the past few years. The study of topological insulators originated with electronic states in condensed matter physics \cite{lu2014topological,hasan2010colloquium,von1986quantized,kane2005quantum}, and were later studied in electromagnetic materials \cite{haldane2008possible,wang2008reflection, khanikaev2013photonic, hafezi2013imaging}. Recently, topological insulators have been investigated for phononic systems (elastic waves in solids) to control statical floppy modes \cite{kane2014topological,paulose2015topological, rocklin2016mechanical, vitelli2014topological, stenull2016topological, bilal2017intrinsically, prodan2017dynamical} and dynamical edge waves \cite{susstrunk2015observation, kariyado2015manipulation}. These mechanical structures are induced by analogs to the quantum Hall \cite{fleury2014sound}, quantum spin Hall \cite{huber2016topological}, or quantum valley Hall effects \cite{lu2016valley}, and inherently support topological edge states resistant to back-scattering at sharp interfaces, discrete defects, or continuous disorder \cite{mousavi2015topologically, miniaci2018experimental}.

Mechanical TIs mimicking the quantum Hall effect (QHE) require active means to break time-reversal symmetry  \cite{khanikaev2015topologically, fleury2014sound,vila2017observation} and have been achieved using a weak magnetic field \cite{prodan2009topological}; gyroscopes or rotating frames \cite{wang2015coriolis, wang2015topological, nash2015topological}; or by varying material properties in time and space \cite{chaunsali2017demonstrating, fleury2016floquet,swinteck2015bulk}. In contrast, mechanical TIs mimicking the quantum spin Hall effect (QSHE)  \cite{mousavi2015topologically, huber2016topological} may be achieved passively through breaking of spatial inversion symmetry. These topological insulators have been investigated numerically, and tested experimentally for elastic waves in thin plates \cite{miniaci2018experimental, chaunsali2018subwavelength, he2016topological}, and in discrete systems composed of masses and linear springs \cite{huber2016topological, prodan2017dynamical, chaunsali2017demonstrating, susstrunk2015observation}. Compared to active TIs, they i) do not require external energy input, ii) feature both forward- and backward-propagating edge modes, and iii) typically retain time-reversal symmetry \cite{landau2013electrodynamics}. 

A third approach for achieving a mechanical TI results from mimicking the quantum valley Hall effect (QVHE), which breaks inversion symmetry in a simpler system. The quantum valley Hall effect was first predicted theoretically in graphene \cite{rycerz2007valley,xiao2007valley,zhang2011spontaneous}, and later observed in solid-state devices \cite{mak2014valley,gorbachev2014detecting,sui2015gate}, photonic crystals \cite{ma2016all,dong2017valley,xiao2015synthetic}, and graphene bi-layers \cite{zhang2013valley,zhang2009direct,ju2015topological}. Unlike QSHE, only one set of degenerate Dirac cones is required, which reduces the geometrical complexity of designing TIs for elastic media. Recently, QVHE has been extended to phononic systems to exhibit valley edge states \cite{lu2016valley} by utilizing: i) anisotropic scatterers in sonic crystals \cite{lu2017observation}, and ii) arrays of resonators or different inclusion types in thin plates \cite{pal2017edge, vila2017observation, darabi2019reconfigurable}. 

Mechanical topological insulators proposed to date lack an easy means of reconfigurabilty, which is essential for enabling important TI-based applications. One potential means to overcome this issue, as explored in this Letter, is to employ shunted piezoelectric disks in which the system's mechanical impedance can be altered dramatically using negative capacitance circuits \cite{trainiti2019time, behrens2003broadband, beck2013power}. Dynamic reconfigurability of such structures can then be obtained through simple ON/OFF switching of these external circuits.  By doing so, we propose and experimentally verify the first electroacoustic topological insulator which exhibits topologically-protected edge states. This reconfigurable structure is composed of an elastic hexagonal lattice (made from polylactic acid, PLA, plastic) whose unit cell contains two shunted piezoelectric (PZT) disks, each connected to a negative capacitance circuit by an ON/OFF switch. Closing one or the other circuit results in the breaking of mirror symmetry and yields mechanical behavior analogous to the QVHE. 

\section{Results}
\subsection{Graphene-like unit cell}
Figure~\ref{fig1}\textbf{(a)} displays the schematic of the unit cell composed of an $h_h=0.5$~$mm$ thick PLA layer, with two bonded piezoelectric disks, used to mimick the band structure of graphene. This unit cell is then periodically repeated in the lattice directions to form the entire material system (see Supplementary Note 1 for more details). Each of the PZT disks employed has a diameter of $7$~$mm$, and a thickness of $h_p=0.5$~$mm$, and is connected to a negative capacitance circuit through a digital ON/OFF switch. Red dashed lines in Fig.~\ref{fig1}\textbf{(b)} plot the band structure of the unit cell when both of the switches are off (i.e., both of PZT disks experience open circuit conditions), documenting two Dirac cones at the edge of the unit cell where two distinct Lamb modes \cite{lamb1917waves} meet (approximately $f_1=45$~$kHz$, $f_2=73$~$kHz$). Note that these curves are computed only for the frequency range of interest; hence, other Dirac cones are not visible. Supplementary Note 1 provides full discussion on the computation details for the entire frequency spectrum ($0-90$~$kHz$).

\subsection{Breaking inversion symmetry}
\begin{figure}
 \centering
\includegraphics[width=3 in]{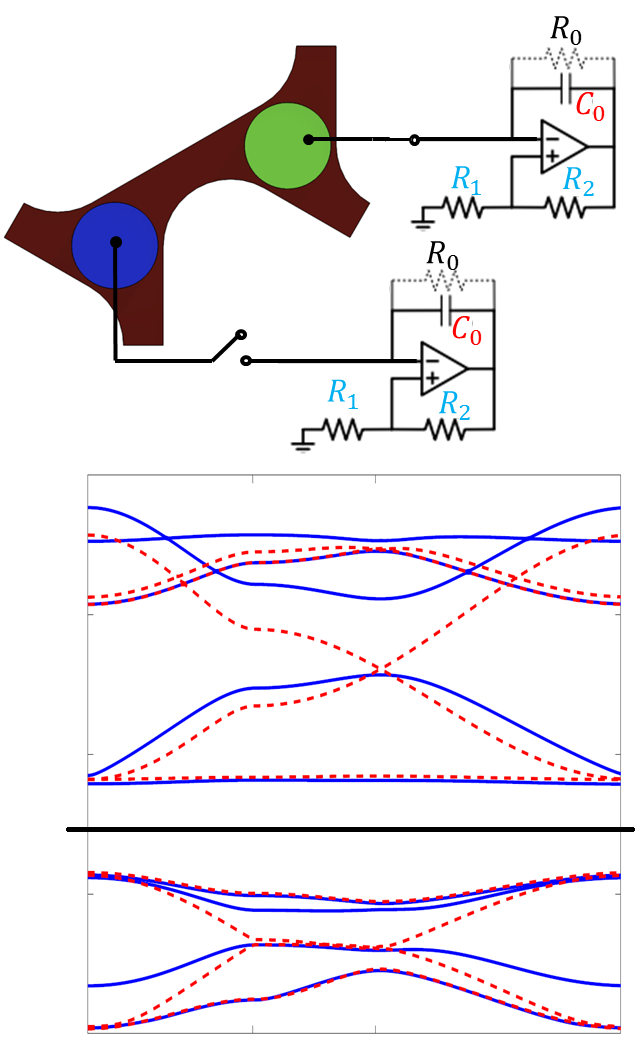}
    \put(-210,320){\textbf{(a)}}
    \put(-210,200){\textbf{(b)}}
  \put(-220,65){\rotatebox{90}{Frequency (kHz)}}
 \put(-125,-18){Wave vector}
\put(-188,-8){$\Gamma$}
 \put(-135,-8){$M$}
 \put(-94,-8){$K$}
 \put(-7,-8){$\Gamma$}
     \put(-198,0){$43$}
  \put(-198,46){$47$}
 \put(-198,94){$70$}
 \put(-198,141){$75$}
 \put(-198,189){$80$}
 \put(-55,120){$C_v=+1/2$}
 \put(-120,140){$C_v=-1/2$}
 \put(-120,55){$C_v=+1/2$}
 \put(-60,30){$C_v=-1/2$}
 \put(-80,318){ON}
 \put(-145,220){OFF} 
     \caption{\textbf{Phononic crystal band structure.} \textbf{(a)} Schematic of the phononic crystal formed by hexagonal unit cells with PLA ($h_0=0.5$~$mm$ thick) as the host layer and attached circular piezoelectric patches with thickness of $h_1=0.5$~$mm$ ($d_1=7$~$mm$ in diameter) connected to external circuits. These circuits provide negative capacitance of $C'=-(R_2 C_0)/R1$. \textbf{(b)} Comparison between the band structures in the absence (red dashed-curves with two Dirac Dirac points at $45$, and $73$~$kHz$), and in the presence (blue solid-curves with optimal bandgap at $C'=-1.7$~$nF$) of completed external circuits attached to one PZT disk.
}\label{fig1}
\end{figure}
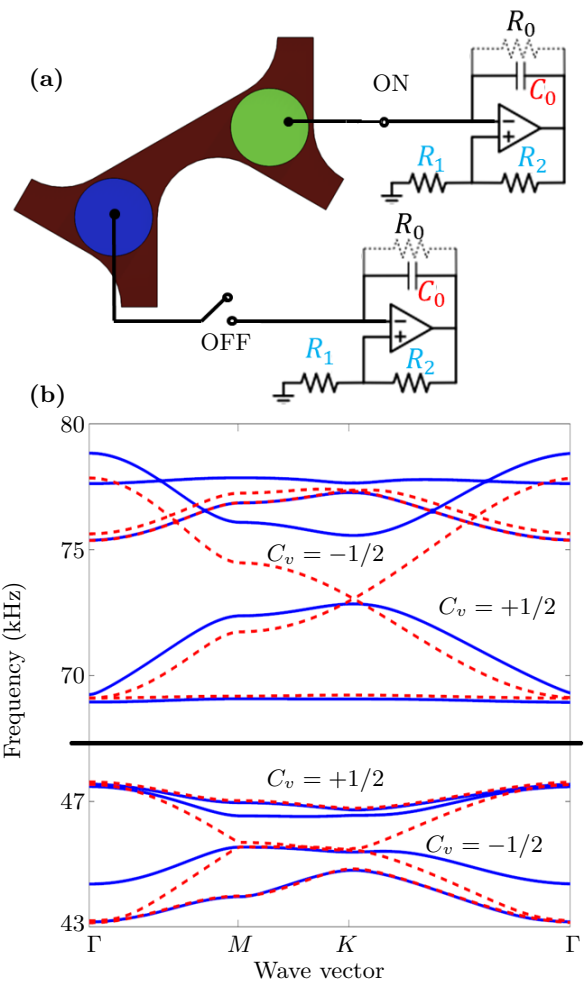

The next step in configuring the TI requires breaking mirror symmetry and separating the folded Dirac cones. As such, one of the switches in Fig.\ref{fig1}\textbf{(b)} is set to ON, thus connecting the PZT (green disk) to the external negative capacitance circuit (Fig.~\ref{fig1}\textbf{(a)}), which provides a significant change in the elastic modulus of the disk \cite{behrens2003broadband, beck2013power, hagood1991damping}; the other PZT (blue disk) remains disconnected.  This then enables breaking $C_6$ symmetry and creates a topological bandgap at the location of the Dirac cone. The circuit includes two resistors ($R_1,R_2$), one capacitor ($C_0$), and an operational amplifier, which yields an effective negative capacitance of $C'=-(R_2C_0)/R_1$. Note that, the resistor $R_0$ prevents saturation of the paralleled capacitance $C_0$, which can cause instability of the PZT disk. Blue solid lines in Fig.~\ref{fig1}(b) plot the band structure of the described unit cell when the green PZT disk is connected in series with a negative capacitance of $C'=-1.7$~$nF$, reporting two complete frequency bandgaps at the location of the Dirac modes. The stated negative capacitance is achieved by placing $R_1=10$,$R_2=17$~$\Omega$, and $C_0=$~$1nF$, resulting in an optimal bandgap width. The band structure for other values of negative capacitance are provided in Supplementary Note 1. 

For each of the bands bounding the topological gaps in Fig.~\ref{fig1}(e), the valley Chern numbers are computed numerically to be $C_v=\pm 1/2$ (as labeled on the graph). The computation of these numbers and the corresponding Berry curvatures are detailed in Supplementary Note 2. According to the bulk-edge correspondence principle \cite{halperin1982quantized,ma2016all}, for each of these bandgaps, the total Chern number is equal to the summation of the Chern numbers ($\Delta C_v$) for all the modes below the gap. If two structures with opposite total Chern numbers for bands share an interface (i.e., one structure with green disk shunted and the other with blue disk shunted), one U-shape helical edge mode will be present at the interface ($|\Delta C_v|=1$). 

\subsection{Topologically protected edge waves} 
\begin{figure*}
 \centering
\includegraphics[width=6 in]{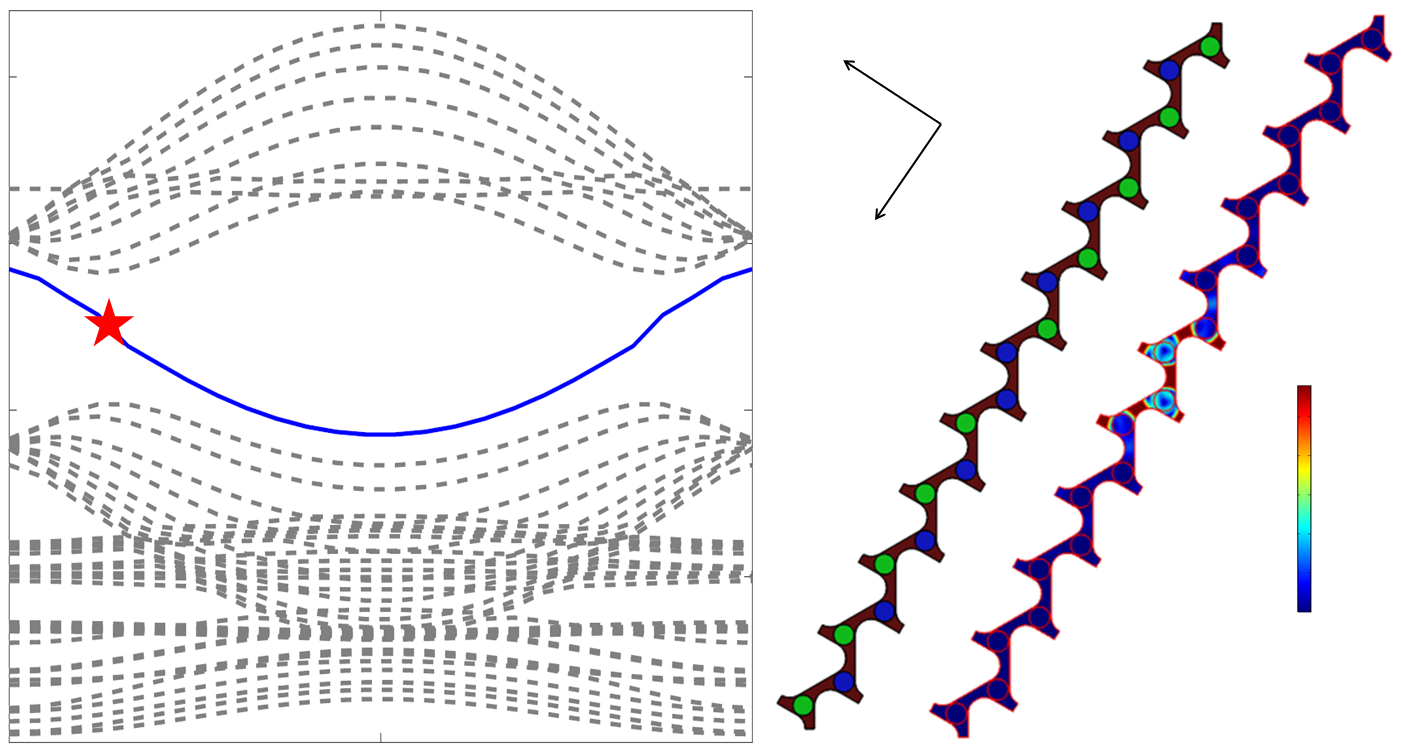}
 \put(-460,70){\rotatebox{90}{Frequency (k Hz)}}
  \put(-465,220){\textbf{(a)}}
  \put(-170,220){\textbf{(b)}}
\put(-443,206){$80$} 
 \put(-443,156){$75$}
  \put(-443,103){$70$}
 \put(-443,51){$65$}
 \put(-443,2){$60$}
 \put(-330,-15){$k_x (\pi/L)$}
\put(-436,-7){$-1$}
 \put(-203,-7){$1$}
 \put(-25,110){$1$}
 \put(-25,40){$0$}
  \put(-175,205){$x$}
 \put(-168,160){$y$}
 \caption{\textbf{Topologically protected interface state.} \textbf{(a)} Band structure of a super-cell composed of a 1-by-10 array of unit cells, periodically repeated in the $x-$direction computed using Floquet boundary conditions. Gray curves depict bulk bands, while the blue curve depicts the protected interface mode. \textbf{(b)} Left: schematic of the super-cell in which green disks are electrically shunted and blue ones are left open circuited, resulting in the upper-half and lower-half subdomains experiencing opposite Chern numbers; Right: the corresponding wave distribution at a frequency of $f=73$~$kHz$ (marked with a red star), documenting wave localization at the interface. All displacements are normalized by the maximum deformation of the cell. 
 }\label{fig2}
\end{figure*}
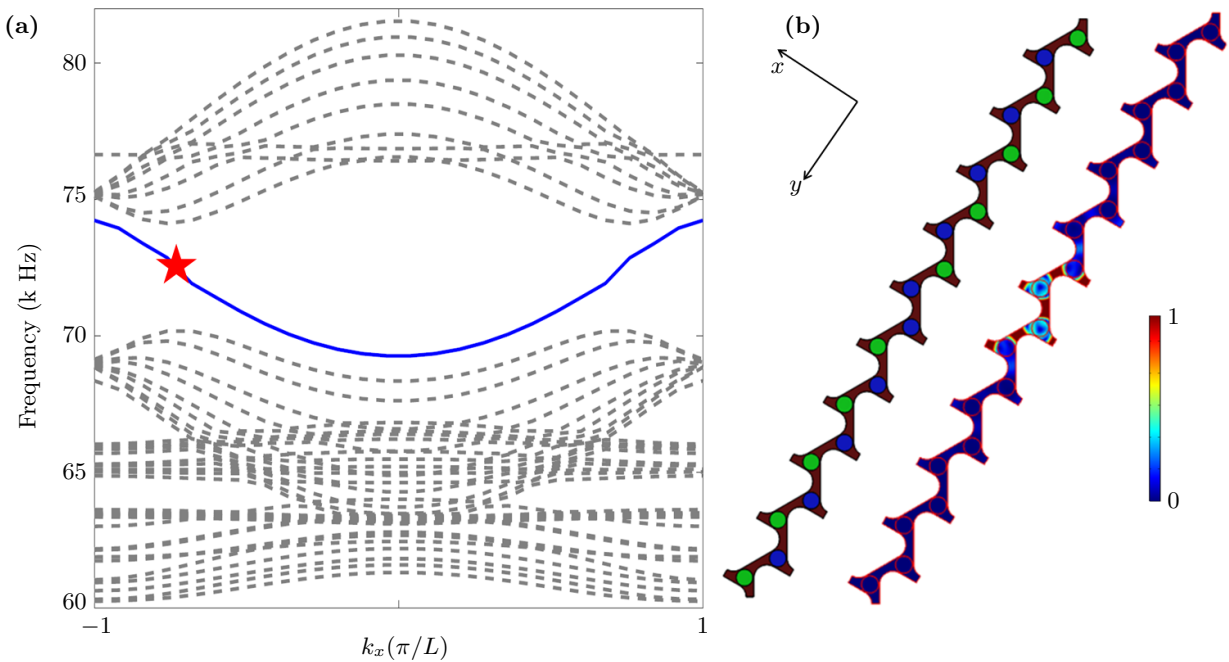

The most intriguing property of topological edge states is their ability to convey waves along sharp and curved interfaces without back-scattering. For QVHE, edge states are not present at the edge of the structure with a trivial mirror (e.g., air) since the difference in Chern numbers is less than one ($|\Delta C_{edge}|=\pm 1/2$); however, according to the bulk-edge correspondence principle \cite{halperin1982quantized, ma2016all}, if two materials with opposite Chern numbers share an interface ($|\Delta C_{interface}|=\pm 1$), topologically protected waves travel along this interface without losing their intensity. To verify these states in our system, a super cell composed of 10 unit cells in the $y-$direction is considered (see Fig.~\ref{fig2}). This depicted strip is then repeated in the $x-$ direction by applying Bloch boundary conditions. As depicted in Fig.~\ref{fig2}(b), all the green disks are electrically shunted, while the blue ones experience an open circuit condition. For the upper half of the super cell, locations of the green disks are reversed compared to the lower half, resulting in an interface with a Chern number difference of $|\Delta C_{interface}|=+1$. Figure \ref{fig2}\textbf{(a)} documents the band structure of the super cell, revealing the existence of a topologically-protected edge state starting from near the bulk modes on top to the bulk modes on the bottom at the location of the upper bandgap in Fig.~\ref{fig1}\textbf{(b)}. Furthermore, the right schematic in Fig.~\ref{fig2}\textbf{(b)} provides the displacement field of the strip at $73$~$kHz$ (marked with a red star). As observed, interface wave is localized at the interface.       

\subsection{Numerical demonstration of edge states} 
Next we report wave propagation immune to back-scattering along desired trajectories. For an ideal topological insulator, data should travel along sharp and curved trajectories without loss of intensity. For mechanical topological insulators, this advantageous behavior has significant implications for communications systems, multiplexers, and on-board mechanical logic. To illustrate such a capability, a reconfigurable interface is created between two sub-domains with opposite Chern numbers. 

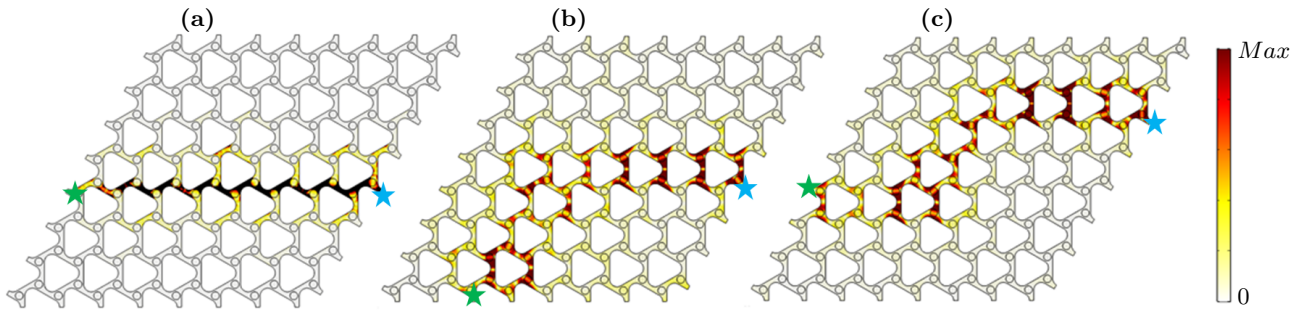
\begin{figure*}
 \centering
\includegraphics[width=6.5 in]{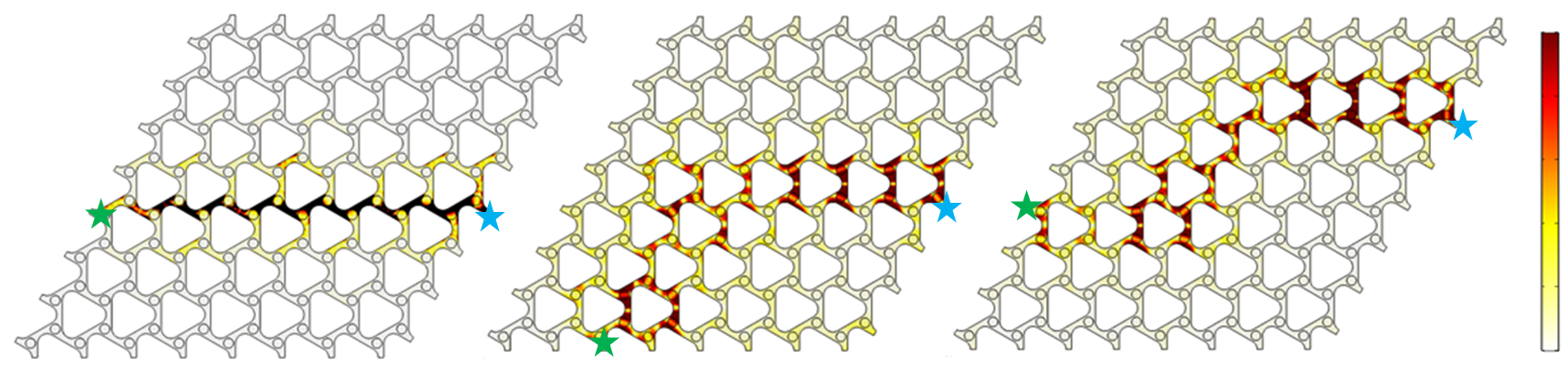}
  \put(-400,110){\textbf{(a)}}
  \put(-260,110){\textbf{(b)}}
  \put(-120,110){\textbf{(c)}}
  \put(0,98){$Max$}
 \put(0,5){$0$}  
 \caption{\textbf{Numerically computed interface waves.} Numerically computed RMS-displacement field of the system excited by a source (marked with blue stars) at $73$~$kHz$, documenting back-scattering free wave propagation along \textbf{(a)} a horizontal interface, \textbf{(b)} a triangular-shaped interface, and \textbf{(b)} a $z$-shaped interface. All displacements are normalized by the amplitude of the input wave. For these interfaces, the location of the shunted piezoelectric patch on either side of the interface is reversed, providing $\Delta C_{interface}=\pm1$ at the boundary of the two sub-domains. 
 }\label{fig3}
\end{figure*}

As such, we consider a plate composed of $8\times8$ hexagonal unit cells, each equipped with a pair of piezoelectric disks. On one side of the interface the upper disks are shunted; while for the other side the lower ones are electrically shunted (or vice-versa).  Figure \ref{fig3}~\textbf{(a)} depicts the displacement field of the structure with a horizontal interface under a harmonic excitation at $73$~$kHz$. As documented, waves clearly travel along the desired interface from the input (marked with the blue star) to the output (marked with the green star). As a second example, Fig.~\ref{fig3}\textbf{(b)} displays an interface in which waves are guided along a triangular path from the source on the left side to the receiver on the bottom edge, without back-scattering or reflection at the sharp edges (again at $73$~$kHz$). Finally, Fig.~\ref{fig3}\textbf{(c)} depicts the displacement field for the case with a $z$-shaped interface, documenting propagation with minimal intensity loss for $73$~$kHz$. As desired, for all three interfaces, robustness of the system is guaranteed at the location of the interface.

\subsection{Experimental realization of edge states}
\begin{figure*}
 \centering
\includegraphics[width=6.5 in]{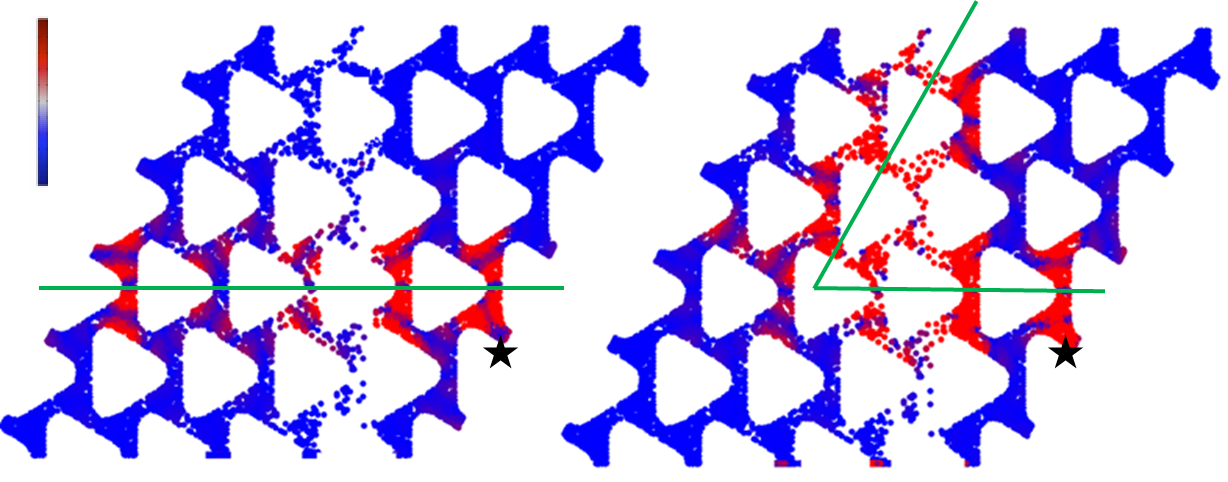}
  \put(-420,170){\textbf{(a)}}
  \put(-200,170){\textbf{(b)}}
  \put(-449,172){$Max$}
  \put(-449,112){$0$}
 \caption{\textbf{Experimentally measured interface waves.} Experimentally-measured RMS displacement field of the system excited by a source (marked with black stars) at $77$~$kHz$, documenting wave propagation \textbf{(a)} along a horizontal interface, and \textbf{(b)} along a sharp-angled interface exhibiting back-scattering free propagation. Displacements are normalized by the amplitude of the input wave. For these interfaces, the location of the shunted piezoelectric patch on either side of the green line is reversed. 
 }\label{fig4}
\end{figure*}

We next verify the performance of the proposed reconfigurable TI by carrying out a set of experiments. The experimental setup is composed of a $5\times 5$ hexagonal array of unit cells incorporating fifty bonded piezoelectric disks (see Supplementary Note 3 for full details on the experimental setup). Half of the PZT disks are connected to external circuits (by simply closing the attached switch) with an effective negative capacitance of $C'=-1.7$~$nF$ to form an interface between two domains (marked with green line), one characterized by a valley Chern number $C_v=1/2$, and the other by $C_v=-1/2$, respectively. Figure \ref{fig4}\textbf{(a)} exhibits the experimentally measured RMS wavefield of a system with a horizontal interface in response to excitation at $77$~$kHz$. This figure clearly confirms the propagation of an interface wave from the source (marked with a black star) to the receiver on the other side of the structure. Due to the imperfectness of the experimental setup, and the mass and stiffness of the connected wires and soldering material, this frequency is slightly above that predicted using numerical simulations. As shown in Fig.~\ref{fig4}(b), the system is easily reconfigured to  alter the interface location (at the same excitation frequency) simply by operation of the programmable ON/OFF switches, in this case introducing a sharp-angled trajectory. An interface wave travels from the input on the left side to the output on the top edge with minimal reflections from the sharp geometry. For both trajectories appearing in Fig.~\ref{fig4}, time snapshots of the interface wave traveling along the trajectories are provided in Supplementary Note 3.

\section{Discussions}
In summary, this Letter presents the first reconfigurable electroacoustic topological insulator to realize topological edge states analogous to the quantum valley Hall effect. The proposed phononic crystal is formed by periodically repeating a graphene-like unit cell, which is composed of two bonded piezoelectric circular disks attached to a hexagonal PLA layer. In order to break inversion symmetry, one of the two disks is connected to an external circuit with negative capacitance, while the other one is left in an open circuit state. Dispersion relationships are computed numerically, documenting a topological bandgap at the location of the Dirac one. Numerically computed and experimentally measured results illustrate immune to back-scattering wave propagation along sharp interfaces. Furthermore, by simply operating ON/OFF the desired circuits, reconfigurable interfaces are obtained and verified experimentally. The reconfigurable TI proposed in this study may be a \textit{stepping-stone} material platform for implementing next generation acoustic-based wave filtering, multiplexing/demultiplexing, and logic in communication-based devices.
 

\bibliography{ref.bib}
\section{Acknowledgments}
The authors would like to thank the National Science Foundation for supporting this research under Grant No. 1929849.

\section{Competing financial interests}
The authors declare no competing financial interests.
\end{document}